\documentstyle[12pt,epsf]{article}
\begin{document}
\begin{titlepage}
\begin{flushright}
ULB--TH 97/21\\
February 1998
\end{flushright}
\vspace*{1.6cm}
\begin{center}
{\Large\bf Central pseudoscalar production in $pp$ scattering and the gluon
contribution to the proton spin}\\
\vspace*{0.8cm}
P.~Castoldi, R.~Escribano\footnote{Chercheur IISN.} and 
J.-M.~Fr\`ere\footnote{Directeur de recherches du FNRS.}\\
\vspace*{0.2cm}
{\footnotesize\it Service du Physique Th\'eorique, Universit\'e Libre de
Bruxelles, CP 225, B-1050 Bruxelles, Belgium}\\
\end{center}
\vspace*{1.2cm}
\begin{abstract}
Central pseudoscalar production in $pp$ scattering is suppressed at small
values of $Q_\perp$, where $Q$ is defined as the difference between the
momenta transferred from the two protons. Such a behaviour is expected if the 
production occurs through the fusion of two vectors. Photon
exchange could provide the dominant contribution at low transferred momenta,
but we argue that an extension of the experiment could probe
the gluon contribution to the proton spin.
\end{abstract}
\end{titlepage}
\section{Introduction}
The use of a glueball-$q\bar q$ filtering method has been recently advocated
to study central hadron production in $pp$ scattering \cite{CLO}. At this
occasion, it was noticed that, somewhat surprisingly, pseudoscalar production
(and in general $q\bar q$ mesons production) was suppressed at small values of
$Q_\perp$ \cite{WA102}, where $Q$ is defined as the difference of the momenta
transferred from the two protons.

We show in this note that such a behaviour is precisely expected if a 
pseudoscalar meson is produced through the fusion of two vector intermediaries.
Furthermore, we argue that an extension of the experiment 
would test the gluon contribution to the proton spin.

In Sect.~2 we introduce the basic formul\ae. In Sect.~3 we try to work a way 
around some important experimental cuts to provide suggestions for comparison
to the data. In particular, we provide the $Q_\perp$ distribution at fixed 
$k_\perp$, $k$ being the momentum of the resonance $X$. This allows for a 
comparison between $\pi^0$, $\eta$ and $\eta\prime$ production and a test of
the nature of the process. Finally, in Sect.~4 we advocate
extending the study to non-exclusive channels 
$pp\rightarrow \tilde p\tilde pX$, where $\tilde p$ are jets corresponding to
$p$ fragmentation, to observe the {\it QCD} equivalent of the process. We then
argue that a measurement of the production cross section at $Q_\perp=0$ would
provide a test of the gluon contribution to the proton spin.
\section{The basic formul\ae}
The WA102 and NA12 experiments \cite{WA102,NA12} have examined in kinematical
detail the reaction $pp\rightarrow ppX$ where $X$ is a single resonance 
produced typically in the central region of the collision between a proton 
beam and an hydrogen target.

We will be more particularly interested in the case where $X$ is a $J^P=0^-$
state, notably $\pi^0$, $\eta$ or $\eta\prime$. The interest of this experiment
is that the kinematics are entirely determined (in fact overdetermined) since
the momenta of all protons are known and the disintegration of $X$ is entirely
measured (e.g. in the $\gamma\gamma$ mode).

The production cross section is affected by two distinct mechanisms: {\it i)}
the emission of the intermediaries from the protons and {\it ii)} the fusion of
those intermediaries into the resonance $X$.

We will mainly interested in the low transferred momenta r\'egime, {\it i.e.}
in the low $t_1$ and $t_2$ region, where $t_1$ and $t_2$ are defined as the
square of the momentum exchanged at the proton vertices. For this reason,
we consider as intermediaries only the lowest-lying particles, mainly
pseudoscalars, vectors and axials. We do not consider heavier particles, in
particular tensors, whose couplings involve a large number of derivatives and
are then expected to contribute less in the kinematical region considered.

In this framework,
the production of a pseudoscalar resonance through the fusion of two
intermediaries in parity conserving interactions could arise from
scalar-pseudo\-scalar $(SP)$ or vector-axial $(VA)$ fusion if no factor of 
momenta is allowed, or, vector-pseudoscalar $(VP)$, vector-vector $(VV)$ or 
axial-axial $(AA)$ fusion if the momentum variables can be used 
\cite{CLO2,ARE}. 
Since the first axial resonance is rather heavy, we also do not consider
the case where one or two of these resonances are involved, so we restrict 
our discussion to $SP$, $VP$ or $VV$ fusion.

In the case of $SP$ fusion, the only pseudoscalar which could be involved in 
the $\pi^0$, $\eta$ and $\eta\prime$ production is the particle itself, but we
still need to find a low-lying scalar, possibly the ``sigma'' or a ``pomeron''
state. Moreover, due to the absence of any derivative coupling, the observed
suppression of the production cross section at small $Q_\perp$ cannot occur 
since non trivial helicity transfer is needed (see Ref.~\cite{ARE} for
details). In the case of $VP$ fusion, the $VPP$ coupling involves one
derivative and should obey Bose and $SU(3)$ symmetry. For instance, a 
$\rho^0\pi^0\pi^0$ coupling is well-known to be forbidden. We conjecture that
the argument can be extended to $U(3)$ symmetry (in particular 
$\rho^0\eta\prime\pi^0$), which removes the discussion of $VP$ fusion from our
analysis.
This leaves $VV$ fusion as the only alternative.

Vector-vector fusion is possible through the vector-vector-pseudoscalar $(VVP)$
coupling
\begin{equation}
\label{VVP}
C_{VVP}=\epsilon_{\mu\nu\alpha\beta} q_1^\mu q_2^\nu
\epsilon_1^\alpha\epsilon_2^\beta\ ,
\end{equation}
where $q_1$ and $q_2$ are the momenta of the exchanged vectors with
polarizations $\epsilon_1$ and $\epsilon_2$ respectively. This coupling is 
well known from the anomalous decay $\pi^0\rightarrow\gamma\gamma$. When 
evaluated in the $X$ rest frame with $k=q_1+q_2$ and $Q=q_1-q_2$, it yields
simply
\begin{equation}
\label{VVPXrf}
C_{VVP}=-\frac{1}{2} m_X \vec{Q}\cdot 
(\vec{\epsilon}_1\times\vec{\epsilon}_2)\ ,
\end{equation}
where clearly the difference $\vec Q$ between $q_1$ and $q_2$ 3-momenta
appears now as a factor and we thus expect a suppression at small $\vec Q$.
But this is insufficient in itself to explain the suppression observed at small
$Q_\perp=|{\vec Q}_\perp|$, where ${\vec Q}_\perp$ is defined as the vector
component of $\vec Q$ transverse to the direction of the initial proton beam.
However, as seen from (\ref{VVPXrf}), the polarizations of the vectors play an
essential role. In particular, in the $X$ rest of frame, 
${\vec\epsilon}_1\times {\vec\epsilon}_2$ must have components in the $\vec Q$
direction, which implies that both ${\vec\epsilon}_1$ and ${\vec\epsilon}_2$
must have components in the plane perpendicular to $\vec Q$, that is, the 
exchanged vectors must have transverse polarization (helicity $h=\pm 1$). In
other terms, the production process will be proportional to the amount of
intermediate vectors with $h=\pm 1$.

If we consider now the emission of a vector from a fermion, we observe that in
the high-energy limit the vector only couples to $\bar f_L\gamma^\mu f_L$ and 
$\bar f_R\gamma^\mu f_R$, that is, the helicity of the fermion cannot change.
In the $X$ rest frame, assumed to lie in the central region of the production,
the colliding fermions cannot (unless they were backscattered, a situation
contrary to the studied kinematical region) emit $h=\pm 1$ vectors in the
forward directions, as this would violate angular momentum conservation.

We thus reach the conclusion that in the above-mentioned kinematical situation,
the production of pseudoscalar mesons by two-vector fusion cannot happen if 
$\vec Q$ is purely longitudinal, but requires 
${\vec Q}_\perp\neq{\vec 0}$ \footnote{There is still a loophole: 
${\vec q}_1$ and ${\vec q}_2$ must have transverse components, but in 
a small area of phase space we could still have
${\vec Q}_\perp=({\vec q}_1-{\vec q}_2)_\perp=\vec 0$. The explicit calculation
below shows this is not significant.}.

It is easy to write down the differential cross section for the central 
production of pseudoscalar resonance $X$ in the reaction $pp\rightarrow ppX$.
Although it may seen daring to treat the $p$ as a pointlike particle in the
process considered, this approximation of the $ppV$ ($V$ any vector) coupling
seems phenomenologically more reasonable than the use of a quark parton model
when strictly exclusive processes are considered (where $p$ fragmentation is
not allowed for).

We use the following notations: $p_1=(E;0,0,p)$ is the beam proton momentum, 
$p_2$ the target proton momentum, $p_3$ the momentum of the outgoing proton 
closest to the beam kinematical area and $p_4$ the momentum of the outgoing 
proton closest to the target kinematical area. The transferred momenta to the 
intermediate vectors are $q_1=p_1-p_3$ and $q_2=p_2-p_4$ respectively, and the
momentum of the resonance $X$ is then defined as 
$k=(W;{\vec k}_\perp,k_\parallel)=q_1+q_2$ with $k^2=m^2_X$ and
$W=\sqrt{m_X^2+k_\perp^2+k_\parallel^2}$. We also define
$Q=(\omega;{\vec Q}_\perp,Q_\parallel)=q_1-q_2$ as the difference between the 
momenta transferred from the two protons.

The differential cross section reads
\begin{equation}
\label{dcs}
\frac{d\sigma}{dQ_\perp dk_\perp dk_\parallel d\varphi}=
\frac{1}{(2\pi)^4}\frac{1}{128 W E p}
\frac{k_\perp Q_\perp}{|(2p-Q_\parallel)(2E-W)-k_\parallel \omega|}
\overline{{|\cal{M}|}^2}\ ,
\end{equation}
where we have choosen as integration variables: $Q_\perp=|{\vec Q}_\perp|$,
$k_\perp=|{\vec k}_\perp|$, $k_\parallel$ and $\varphi$ defined as the angle
between the two transverse vectors ${\vec k}_\perp$ and ${\vec Q}_\perp$.
The averaged square of the invariant matrix amplitude $\cal{M}$ at the lowest
order in the vector exchange is expressed in terms of kinematical invariants
as\footnote{
Given the experimental condition, the diagram where the ``fast'' and ``slow''
protons are exchanged gives a negligible contribution, and it is not included
in (\ref{M2}).}
\begin{eqnarray}
\label{M2}
\overline{{|\cal{M}|}^2}&=&
\frac{(g_{ppV_1} g_{ppV_2} g_{V_1V_2P})^2}
{(t_1-m_{V_1}^2)^2 (t_2-m_{V_2}^2)^2}\{
\frac{t_1t_2}{2}[(k\cdot Q)^2-m_X^2 Q^2]\nonumber \\[1ex]
&+&t_1[m^2((k\cdot Q)^2-m_X^2 Q^2)+m_X^2(p_2\cdot Q)^2+Q^2(p_2\cdot k)^2
\nonumber \\[1ex]  
& &-2(p_2\cdot k)(p_2\cdot Q)(k\cdot Q)]\\[1ex] 
&+&t_2[m^2((k\cdot Q)^2-m_X^2 Q^2)+m_X^2(p_1\cdot Q)^2+Q^2(p_1\cdot k)^2
\nonumber \\[1ex]  
& &-2(p_1\cdot k)(p_1\cdot Q)(k\cdot Q)]\nonumber \\[1ex] 
&+&4\left(\epsilon_{\mu\nu\alpha\beta} p_1^\mu p_2^\nu
k^\alpha Q^\beta\right)^2\}\ ,\nonumber 
\end{eqnarray}
where $g_{VVP}$ stands for the coupling constant of the $VVP$ interaction,
$g_{ppV}$ stands for the coupling constant of the $ppV$ interaction, 
$t_{1,2}$ are the square of the momentum 
transfer to each vector, $m_V$ is the mass of the exchanged vector, 
$m_X$ the resonance mass and $m$ the proton mass.
In order to write (\ref{dcs}) in terms of the integration variables we have 
used the following substitutions: 
$(p_{1,2}\cdot k)=E W\mp p k_\parallel$,
$(p_{1,2}\cdot Q)=E \omega\mp p Q_\parallel$, 
$(k\cdot Q)=W\omega -k_\parallel Q_\parallel -k_\perp Q_\perp\cos\varphi=
-2p k_\parallel +2E\omega$,
$Q^2=\omega^2-Q_\perp^2-Q_\parallel^2=-m_X^2+4E W-4p Q_\parallel$ and
$t_{1,2}=q^2_{1,2}=E(W\pm\omega)-p(Q_\parallel\pm k_\parallel)$.

Due to the smallness of $t_1$ and $t_2$ we have checked that independently of
the mass of the vector exchanged, the dominant contribution of (\ref{M2}) to
the cross section (\ref{dcs}) comes from the term proportional to 
$1/(t_1-m_{V_1}^2)^2(t_2-m_{V_2}^2)^2$. 
The differential cross section thus simplifies to
\begin{eqnarray}
\label{dcsaprox}
\frac{d\sigma}{dQ_\perp dk_\perp dk_\parallel d\varphi}&\simeq&
\frac{1}{(2\pi)^4}\frac{1}{128 W E p}
\frac{k_\perp Q_\perp}{|(2p-Q_\parallel)(2E-W)-k_\parallel \omega|}
\nonumber \\[1ex]
&\times&16(g_{ppV_1} g_{ppV_2} g_{V_1V_2P})^2 
E^2 p^2\frac{k_\perp^2 Q_\perp^2\sin^2\varphi}
{(t_1-m_{V_1}^2)^2(t_2-m_{V_2}^2)^2}\ ,
\end{eqnarray}
where now one can clearly see the suppression of the cross section at small
$Q_\perp$ (and indeed $\lim_{Q_\perp\rightarrow 0}d\sigma=0$), as it is seen
experimentally.

Once the expression for the differential cross section is presented, we may
now enter into conjectures about the nature of the vectors exchanged. The 
simplest candidates for elementary particles are of course photon or gluon,
with the possible addition as an example of the massive vectors 
$\rho$, $\omega$ and $\phi$. We will first consider the case of 
$t_1, t_2\rightarrow 0$; it is then quite clear that the 
dominant contribution to the simplified cross section (\ref{dcsaprox}) comes
from the exchange of massless vectors, so we neglect temporarly the
possible contributions of massive vectors. Then,
we are left with photons or gluons. However, in the present situation, gluon
exchange seems not to be the dominant contribution, as it would lead to a 
large number of $\eta\prime$ and $\eta$ and no $\pi^0$, which is clearly not
the experimental situation \cite{WA102ref2}. 
Most probably, the selection of isolated protons
in the final state is too restrictive for gluon exchange to take place
significantly. So then, we conclude that a pure photoproduction hypothesis
may be the main contribution to the cross section at very 
low transferred momenta.

Assuming the photoproduction mechanism as the main effect responsible of the 
pseudoscalar production, we would like to point out that very relevant
information can be obtained here of the $(t_1,t_2)$ behaviour of the
$\gamma\gamma$-pseudoscalar form factor, a question highly discussed in the
literature \cite{GER}.

We will see however that the experiment does not allow isolation of this
low $t_1$ and $t_2$ kinematical region, and that at least the lowest vectors
need to be included.
\section{Working around the cuts}
As we have seen in the previous section, photon-photon fusion could be the
main mechanism responsible for the pseudoscalar production in the central
region for low values of exchanged momenta. 
Thus, we observe that the cross section tends to peak
sharply at small values of $Q_\perp\simeq k_\perp$ (through the photon
denominators), even though it decreases to zero when $Q_\perp$ or 
$k_\perp\rightarrow 0$.

This behaviour is however difficult to observe experimentally, due to the 
presence of experimental set-up restrictions, leading to a loss of 
acceptance when the transverse momenta of the outgoing protons decreases.
This seems to be specially sensitive for the ``slow proton''. As a result,
although the theoretical cross section peaks for small non-zero values of
$Q_\perp\simeq k_\perp$, this domain of parameter space is totally inadequate
for a detailed comparison to experiment.

In practice, we could work at fixed $k_\perp$
in order to avoid the experimental restrictions, and 
explore the $Q_\perp$ dependence of the cross section. In that case,
pseudoscalar production by photon-photon fusion (and in general by any
vector-vector fusion) will be characterized by the vanishing of the cross
section at small $Q_\perp$. Other vector exchanges provide largely enhanced
contributions, retaining the low $Q_\perp$ suppression and with a high
$Q_\perp$ sensitiviness to the form factors used.

We give as examples, curves\footnote{
For simplicity, we have restricted ourselves to the case $k_\parallel=0$.} of
the pseudoscalar photoproduction cross section of 
$\pi^0$, $\eta$ and $\eta\prime$
as a function of $Q_\perp$ for two characteristic values of the angle 
$\varphi$ between ${\vec k}_\perp$ and ${\vec Q}_\perp$.
For $\varphi=45^\circ$ (or in fact, any value of the angle $\varphi$ far
away from $0^\circ$ and $180^\circ$), both $q_{1\perp}$ and $q_{2\perp}$ are
far from zero, and we do not expect acceptance restrictions. On the contrary,
for $\varphi\simeq 10^\circ$ (or in fact, any value of the angle $\varphi$
close to $0^\circ$ and $180^\circ$), the cross section turns out to be much
larger, but we expect here acceptance limitations, since close to 
$Q_\perp=k_\perp$ we can have either $q_{1\perp}\approx 0$ or 
$q_{2\perp}\approx 0$.

While these curves are given as indicatives and seem to have some bearing to
published data, a detailed comparison, which can only be made with an
extensive simulation of the experimental detection, is clearly to be left
to the experimental collaborations.

A work equivalent to the former can be done if instead of using $Q_\perp$ as
the distribution variable for the curves one uses the azimutal angle $\phi$
(not to be confused with the angle $\varphi$ introduced before) between the
fast and slow proton that is related to $Q_\perp$ by 
$\cos\phi=\frac{Q_\perp^2-k_\perp^2}{4q_{1\perp}q_{2\perp}}$.

The above considerations allowed us to compute completely a simplified
cross section. However the experimental difficulty to reach very small values
of $t_1$ and $t_2$ invalidates largely the assumptions made. In this case
indeed, similar contributions will arise from vector bosons, and their much
stronger couplings will offset the mass supression in the propagator.

We have performed such a calculation using formula (\ref{dcsaprox}) above.
The $VVP$ coupling coefficients can be obtained along the lines of \cite{BALL}
and are given in Tables \ref{table1}--\ref{table3},
while the $ppV$ ones are estimated in \cite{NAG}.

The behaviour obtained confirms the low $Q_\perp$ suppression, but the 
general structure of the curve and its peak value are very sensitive to 
vertex form factors, on which we have little independent information.
These form factors, (both at the proton and pseudoscalar vertex) can be 
combined in a single function $f(t_1)\cdot f(t_2)$, which could of course be
fitted directly from experiment, much as the traditional $\exp(-b t)$ usually
is. 

This offers on one hand the possibility to gather information
on form factors, in particular on the $VVP$ ones \cite{GER2}, but as the
main point of the paper is concerned (Sec.~4 below), this ``background''
does not affect the conclusions (since only the $Q_\perp\rightarrow 0$
suppression is of importance).
\section{Extending the approach to gluons}
In this final section, we would like to advocate for an extension of the 
present study to non-exclusive processes $pp\rightarrow\tilde p\tilde pX$,
where $\tilde p$ are jets corresponding to $p$ fragmentation, in order to
observe the {\it QCD} equivalent of the production mechanism 
(gluon-gluon fusion).

In this case indeed, we must distinguish between gluons emitted from the
fermionic partons (and obeying the helicity constraints discussed at the
beginning of the previous section) and ``constituents'' or ``sea'' gluons.
The latter simply share part of the proton momentum and their helicity is in
no way constrained. Helicity $h=\pm 1$ gluons can then be met even for 
${\vec Q}_\perp=\vec 0$, and in that case we would expect that the production
distributions in $Q_\perp$ could be considerably affected.

In this possible extension of the experiments, the $\eta\prime$ and $\eta$ now
produced at small $Q_\perp$ are sensitive to the polarization of the 
individual gluons in the proton. Such polarization of the individual gluons is
always present independently of the total polarization of the gluons in the
proton, and is in itself not indicative of the fact that a significant 
proportion of the proton spin could be carried by the gluons. 
If such would be
the case however, and a net polarization of the gluons exists, a similar 
experiment conducted with polarized beams or target would 
lead to a difference in the production rates of $\eta\prime$
and $\eta$ at small $Q_\perp$, and provide a direct measurement of this
polarization.

In summary, we have shown in this letter that the experimental evidence of the
suppression at small $Q_\perp$ of the central pseudoscalar production in
$pp$ scattering can be explained if the production mechanism is through the
fusion of two vectors.
We also have proposed an extension of such experiments in order to observe the
{\it QCD} equivalent of the process and to provide a test for the gluon
contribution to the proton spin.

\section{Acknowledgements}
It is a pleasure to acknowledge here the stimulating discussions 
with our friends from the WA102 collaboration, in particular, Freddy Binon, 
Andrew Kirk, Sasha Singovski and Jean-Pierre Stroot, as well as Frank Close
and Pierre Marage.

\clearpage
\begin{figure}
\[
\epsffile{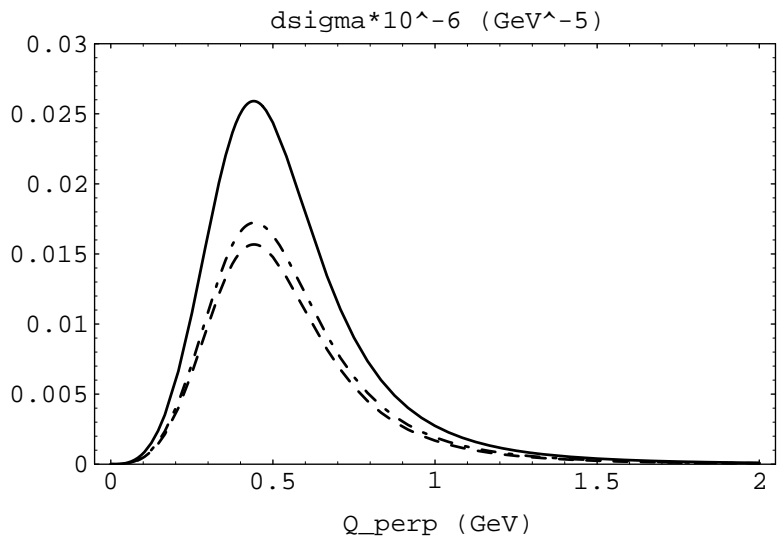}
\]
\vspace*{-2.0cm}
\caption{Differential cross section for $pp\rightarrow ppX$ 
($d\sigma\times 10^{-6}$ (GeV$^{-5}$)), as a function of $Q_\perp$ (GeV). 
We have fixed $k_\perp=0.5$ GeV and $k_\parallel=0$, and this
plot is given for $\varphi=45^\circ$. The curves describe $X=\pi^0$ (solid
line), $X=\eta$ (dashed line) and $X=\eta\prime$ (dashed-dotted line).}
\label{fig1}
\end{figure}

\begin{figure}
\[
\epsffile{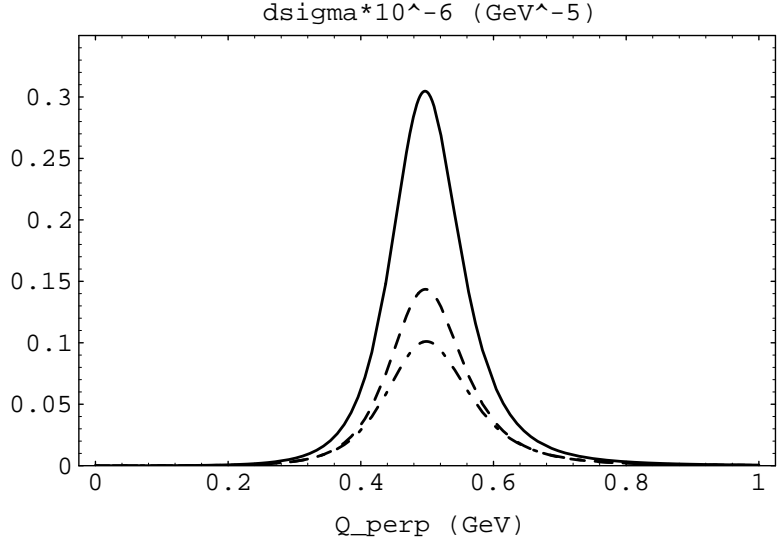}
\]
\vspace*{-2.0cm}
\caption{The same as Fig.~\protect\ref{fig1} but for $\varphi=10^\circ$.}
\label{fig2}
\end{figure}

\begin{table}
\begin{tabular}{lll}\\[2ex] \hline \\[0.1ex]
$\gamma\gamma P$ & $g_{\gamma\gamma P}$ & \\[2ex] \hline \\[0.1ex]
$\gamma\gamma\pi^0$ & $g_{\gamma\gamma\pi^0}\equiv
\frac{\sqrt{2}}{4\pi^2 f_\pi}$ & 
$\simeq 0.27\ \mbox{GeV}^{-1}$ \\[2ex]
$\gamma\gamma\eta$ & $g_{\gamma\gamma\pi^0}
\left(\frac{c\theta}{\sqrt{3}}\frac{f_\pi}{f_8}-
2\sqrt{\frac{2}{3}}s\theta \frac{f_\pi}{f_0}\right)$ & 
$\simeq 0.26\ \mbox{GeV}^{-1}$ \\[2ex]
$\gamma\gamma\eta\prime$ & $g_{\gamma\gamma\pi^0}
\left(\frac{s\theta}{\sqrt{3}}\frac{f_\pi}{f_8}+
2\sqrt{\frac{2}{3}}c\theta \frac{f_\pi}{f_0}\right)$ & 
$\simeq 0.34\ \mbox{GeV}^{-1}$ \\[2ex] \hline
\end{tabular}
\caption{$\gamma\gamma P$ couplings. We have used the following
values: the pion decay constant $f_\pi$=0.132 GeV, the $\eta$-$\eta\prime$
mixing angle $\theta=-16.5^\circ$ and the $\eta_8$ and $\eta_0$ decay constants
$f_8=1.04 f_\pi$ and $f_0=1.12 f_\pi$.}
\label{table1}
\end{table}

\begin{table}
\begin{tabular}{lll}\\[2ex] \hline \\[0.1ex]
$V\gamma P$ & $g_{V\gamma P}$ & \\[2ex] \hline \\[0.1ex]
$\rho^0\gamma\pi^0$ & $g_{\rho^0\gamma\pi^0}\equiv\frac{g}{4\pi^2 f_\pi}$ & 
$\simeq 0.82\ \mbox{GeV}^{-1}$ \\[2ex]
$\omega\gamma\pi^0$ & $g_{\rho^0\gamma\pi^0} 3 c\varphi_V$ &
$\simeq 2.46\ \mbox{GeV}^{-1}$ \\[2ex]
$\phi\gamma\pi^0$ & $g_{\rho^0\gamma\pi^0} 3 s\varphi_V$ &
$\simeq 0.15\ \mbox{GeV}^{-1}$ \\[2ex]
$\rho^0\gamma\eta$ & $g_{\rho^0\gamma\pi^0}\left(
\sqrt{3}c\theta \frac{f_\pi}{f_8}-
\sqrt{6}s\theta \frac{f_\pi}{f_0}\right)$ &
$\simeq 1.82\ \mbox{GeV}^{-1}$ \\[2ex]
$\rho^0\gamma\eta\prime$ & $g_{\rho^0\gamma\pi^0}\left(
\sqrt{3}s\theta \frac{f_\pi}{f_8}+
\sqrt{6}c\theta \frac{f_\pi}{f_0}\right)$ &
$\simeq 1.33\ \mbox{GeV}^{-1}$ \\[2ex]
$\omega\gamma\eta$ & $g_{\rho^0\gamma\pi^0}\left[
c\theta \frac{f_\pi}{f_8}
\left(\frac{c\varphi_V}{\sqrt{3}}-2\sqrt{\frac{2}{3}}s\varphi_V\right)-
s\theta \frac{f_\pi}{f_0}
\left(\sqrt{\frac{2}{3}}c\varphi_V+\frac{2}{\sqrt{3}}s\varphi_V\right)
\right]$ &
$\simeq 0.55\ \mbox{GeV}^{-1}$ \\[2ex]
$\omega\gamma\eta\prime$ & $g_{\rho^0\gamma\pi^0}\left[
s\theta \frac{f_\pi}{f_8}
\left(\frac{c\varphi_V}{\sqrt{3}}-2\sqrt{\frac{2}{3}}s\varphi_V\right)+
c\theta \frac{f_\pi}{f_0}
\left(\sqrt{\frac{2}{3}}c\varphi_V+\frac{2}{\sqrt{3}}s\varphi_V\right)
\right]$ &
$\simeq 0.51\ \mbox{GeV}^{-1}$ \\[2ex]
$\phi\gamma\eta$ & $g_{\rho^0\gamma\pi^0}\left[
c\theta \frac{f_\pi}{f_8}
\left(\frac{s\varphi_V}{\sqrt{3}}+2\sqrt{\frac{2}{3}}c\varphi_V\right)-
s\theta \frac{f_\pi}{f_0}
\left(\sqrt{\frac{2}{3}}s\varphi_V-\frac{2}{\sqrt{3}}c\varphi_V\right)
\right]$ &
$\simeq 1.03\ \mbox{GeV}^{-1}$ \\[2ex]
$\phi\gamma\eta\prime$ & $g_{\rho^0\gamma\pi^0}\left[
s\theta \frac{f_\pi}{f_8}
\left(\frac{s\varphi_V}{\sqrt{3}}+2\sqrt{\frac{2}{3}}c\varphi_V\right)+
c\theta \frac{f_\pi}{f_0}
\left(\sqrt{\frac{2}{3}}s\varphi_V-\frac{2}{\sqrt{3}}c\varphi_V\right)
\right]$ &
$\simeq -1.15\ \mbox{GeV}^{-1}$ \\[2ex] \hline
\end{tabular}
\caption{$V\gamma P$ couplings. We have used the following values:
the $SU(3)$ symmetric coupling constant $g=4.28$ determined by neutral
$\rho$ decay, the departure from $\omega$-$\phi$ ideal mixing parametrized by
the angle $\varphi_V=3.4^\circ$, as well as the values introduced in 
Table \protect\ref{table1}.}
\label{table2}
\end{table}

\begin{table}
\begin{tabular}{lll}\\[2ex] \hline \\[0.1ex]
$VVP$ & $g_{VVP}$ & \\[2ex] \hline \\[0.1ex]
$\rho^0\omega\pi^0$ & $\frac{3\sqrt{2}g^2}{4\pi^2 f_\pi}c\varphi_V$ &
$\simeq 14.86\ \mbox{GeV}^{-1}$ \\[2ex]
$\rho^0\phi\pi^0$ & $\frac{3\sqrt{2}g^2}{4\pi^2 f_\pi}s\varphi_V$ &
$\simeq 0.88\ \mbox{GeV}^{-1}$ \\[2ex]
$\rho^0\rho^0\eta$ & $\frac{6\sqrt{2}g^2}{4\pi^2 f_\pi}\left(
\frac{c\theta}{2\sqrt{3}}\frac{f_\pi}{f_8}-
\frac{s\theta}{\sqrt{6}}\frac{f_\pi}{f_0}\right)$ &
$\simeq 11.00\ \mbox{GeV}^{-1}$ \\[2ex]
$\rho^0\rho^0\eta\prime$ & $\frac{6\sqrt{2}g^2}{4\pi^2 f_\pi}\left(
\frac{s\theta}{2\sqrt{3}}\frac{f_\pi}{f_8}+
\frac{c\theta}{\sqrt{6}}\frac{f_\pi}{f_0}\right)$ &
$\simeq 8.06\ \mbox{GeV}^{-1}$ \\[2ex]
$\omega\omega\eta$ & $\frac{6\sqrt{2}g^2}{4\pi^2 f_\pi}\left[
c\theta\frac{f_\pi}{f_8}\left(
\frac{c^2\varphi_V}{2\sqrt{3}}-\frac{s^2\varphi_V}{\sqrt{3}}\right)-
\frac{s\theta}{\sqrt{6}}\frac{f_\pi}{f_0}\right]$ &
$\simeq 10.92\ \mbox{GeV}^{-1}$ \\[2ex]
$\omega\omega\eta\prime$ & $\frac{6\sqrt{2}g^2}{4\pi^2 f_\pi}\left[
s\theta\frac{f_\pi}{f_8}\left(
\frac{c^2\varphi_V}{2\sqrt{3}}-\frac{s^2\varphi_V}{\sqrt{3}}\right)+
\frac{c\theta}{\sqrt{6}}\frac{f_\pi}{f_0}\right]$ &
$\simeq 8.08\ \mbox{GeV}^{-1}$ \\[2ex]
$\phi\phi\eta$ & $\frac{6\sqrt{2}g^2}{4\pi^2 f_\pi}\left[
c\theta\frac{f_\pi}{f_8}\left(
\frac{s^2\varphi_V}{2\sqrt{3}}-\frac{c^2\varphi_V}{\sqrt{3}}\right)-
\frac{s\theta}{\sqrt{6}}\frac{f_\pi}{f_0}\right]$ &
$\simeq -12.68\ \mbox{GeV}^{-1}$ \\[2ex]
$\phi\phi\eta\prime$ & $\frac{6\sqrt{2}g^2}{4\pi^2 f_\pi}\left[
s\theta\frac{f_\pi}{f_8}\left(
\frac{s^2\varphi_V}{2\sqrt{3}}-\frac{c^2\varphi_V}{\sqrt{3}}\right)+
\frac{c\theta}{\sqrt{6}}\frac{f_\pi}{f_0}\right]$ &
$\simeq 15.07\ \mbox{GeV}^{-1}$ \\[2ex]
$\omega\phi\eta$ & $\frac{3\sqrt{2}g^2}{4\pi^2 f_\pi}
c\theta\frac{f_\pi}{f_8}\frac{\sqrt{3}}{2}c\varphi_V s\varphi_V$ &
$\simeq 0.70\ \mbox{GeV}^{-1}$ \\[2ex]
$\omega\phi\eta\prime$ & $\frac{3\sqrt{2}g^2}{4\pi^2 f_\pi}
s\theta\frac{f_\pi}{f_8}\frac{\sqrt{3}}{2}c\varphi_V s\varphi_V$ &
$\simeq -0.21\ \mbox{GeV}^{-1}$ \\[2ex] \hline
\end{tabular}
\caption{$VVP$ couplings. We have used the values introduced in
Tables \protect\ref{table1} and \protect\ref{table2}.}
\label{table3}
\end{table}

\begin{thebibliography}{99}
\bibitem{CLO} F.~E.~Close and A.~Kirk, Phys. Lett. B 397 (1997) 333.
\bibitem{WA102} D.~Barberis et al., WA102 Collaboration, Phys. Lett. B 397
(1997) 339.
\bibitem{NA12} D.~Alde et al., GAMS Collaboration, Phys. Lett. B 397 (1997)
350.
\bibitem{CLO2} F.~E.~Close, hep-ph/9710450.
\bibitem{ARE} T.~Arens, O.~Nachtmann, M.~Diehl and P.~V.~Landshoff,
Z. Phys. C 74 (1997) 651.
\bibitem{WA102ref2} D.~Barberis et al., WA102 Collaboration, in preparation.
\bibitem{GER} J.-M.~G\'erard and T.~Lahna, Phys. Lett. B 356 (1995) 381.
\bibitem{BALL} P.~Ball, J.-M.~Fr\`ere and M.~Tytgat, 
Phys. Lett. B 365 (1996) 367.
\bibitem{NAG} M.~M.~Nagels et al., Nucl. Phys. B 109 (1976) 1.
\bibitem{GER2} J.-M.~G\'erard and G.~L\'opez Castro, hep-ph/9709404.
\end{thebibliography}
\end{document}